\documentclass[print,showpacs,preprintnumbers,amsmath,amssymb]{revtex4-1}
\usepackage{amsmath}
\usepackage{natbib}
\usepackage{amssymb}
\usepackage{color}
\usepackage{graphicx}
\usepackage{dcolumn}
\usepackage{bm} 
\usepackage{hyperref}
\usepackage{longtable}
\usepackage{gensymb}
\usepackage{siunitx}
\usepackage[version=4]{mhchem}
\usepackage{tabularx}
\usepackage{booktabs}
\newcolumntype{C}[1]{>{\centering\arraybackslash}p{#1}}
\usepackage[section]{placeins}

\begin{document}
\texttt{}
\title{Single spin-polarized Fermi surface in SrTiO$_3$ thin films}

\author{Eduardo~B.~Guedes$^{1,2}$}
\thanks{These authors contributed equally}
\author{Stefan~Muff$^{1,2}$}
\thanks{These authors contributed equally}
\author{Mauro~Fanciulli$^{1,2}$}
\author{Andrew~P.~Weber$^{1,2}$}
\author{Marco~Caputo$^{1,2}$}
\author{Zhiming~Wang$^{3}$}
\author{Nicholas~C.~Plumb$^{2}$}
\author{Milan~Radovi\'{c}$^{2}$}
\author{J.~Hugo~Dil$^{1,2}$}

\affiliation{
\\ 
$^{1}$Institut de Physique, \'{E}cole Polytechnique F\'{e}d\'{e}rale de Lausanne, CH-1015 Lausanne, Switzerland
\\ 
$^{2}$Photon Science Division, Paul Scherrer Institut, CH-5232 Villigen, Switzerland
\\
$^{3}$Key Laboratory of Magnetic Materials and Devices, Ningbo Institute of Materials Technology and Engineering, Chinese Academy of Sciences, Ningbo 315201, Peoples Republic of China
\\
}

\date{\today}

\begin{abstract} 
The 2D electron gas (2DEG) formed at the surface of SrTiO$_3$(001) has attracted great interest because of its fascinating physical properties and potential as a novel electronic platform, but up to now has eluded a comprehensible way to tune its properties. Using angle-resolved photoemission spectroscopy with and without spin detection we here show that the band filling can be controlled by growing thin SrTiO$_3$ films on Nb doped SrTiO$_3$(001) substrates. This results in a single spin-polarised 2D Fermi surface, which bears potential as platform for Majorana physics. Based on our results it can furthermore be concluded that the 2DEG does not extend more than 2 unit cells into the film and that its properties depend on the amount of SrO$_x$ at the surface and possibly the dielectric response of the system. 

\end{abstract}

\maketitle

\section{Introduction}

Transition metal oxides are expected to play an important role in next generation electronics and devices, primarily driven by the interplay of lattice, charge, orbit, and spin degrees of freedom in these materials in combination with correlation effects \cite{Spaldin_review:2010}. A prominent sub-class are the titanates with a perovskite structure (ATiO$_3$) which, despite having a large band gap in the bulk, at interfaces with other materials (or vacuum), develops a high mobility two-dimensional electron gas (2DEG) with a wide variety of intriguing properties \cite{Reyren:2007,Bert:2011,Ariando:2011,Kalisky:2012}. Especially SrTiO$_3$ has been extensively studied, partly because it was the first system where such a 2DEG was observed \cite{ohtomo:2004} and partly because it is an easily accessible material.

\begin{figure}[htb]
	\includegraphics[width=0.95\columnwidth]{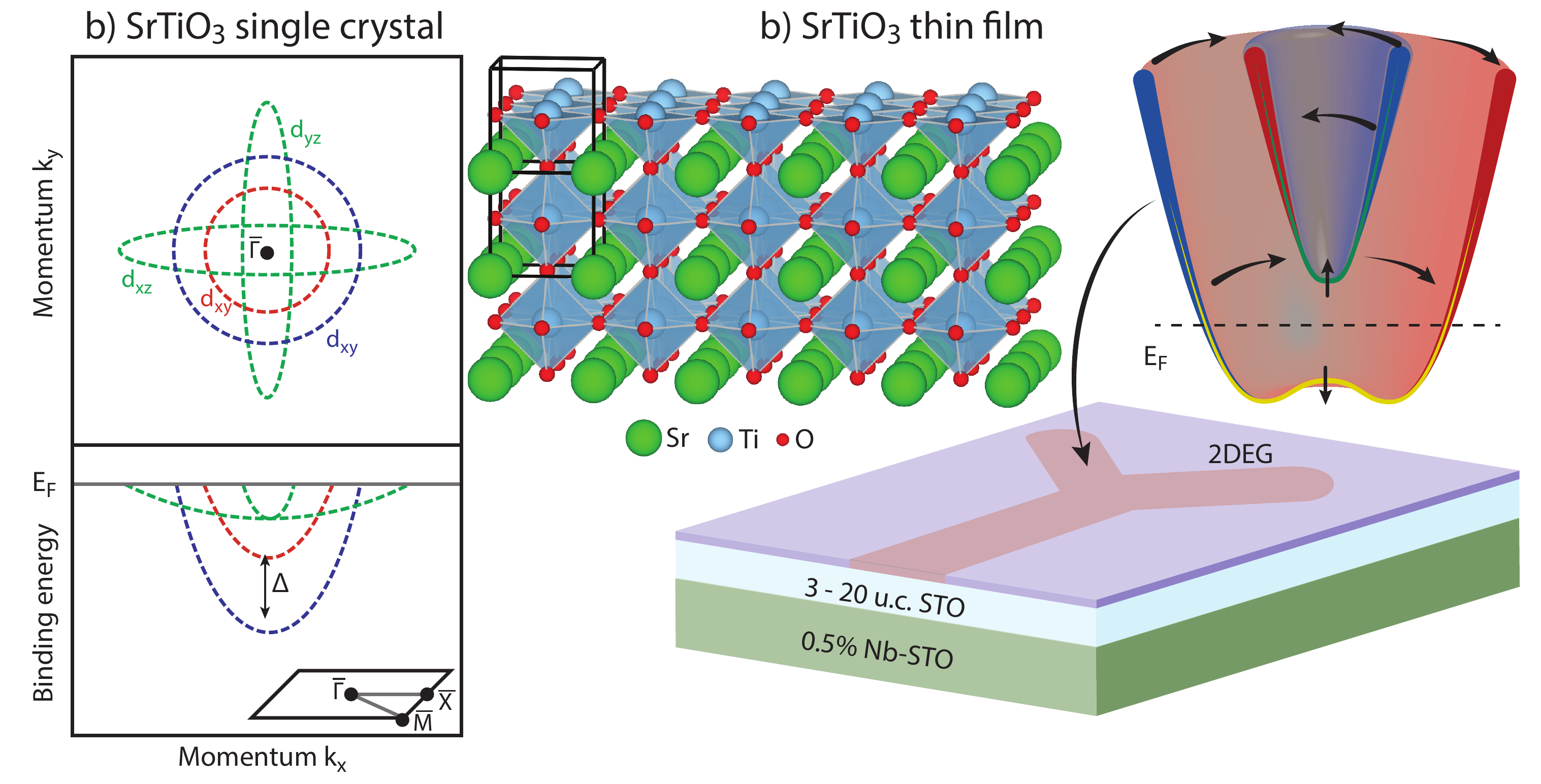}
\caption{The spin-polarised 2DEG on SrTiO$_3$ crystals and thin films. (a) Schematics of the electronic structure of the 2DEG on STO crystals, consisting of two circular $3d_{xy}$-derived states (red and blue) and ellipsoidal $3d_{xz}$- and $3d_{yz}$-derived states (green). The inset shows cubic the surface Brillouin Zone. (b) Illustration showing the crystal structure \cite{Momma:2011} of STO and the STO film grown on 0.5 wt\% Nb-doped STO substrate, along with the band structure of the hosted 2DEG. In this case, the Fermi level lies in the Zeeman gap at the SBZ centre, thus resulting in a single spin-polarised 2D Fermi surface. This band structure, along with the superconducting properties of STO, makes this material a 2D Majorana platform.}
	\label{fig.1}
\end{figure}

Angle-resolved photoemission spectroscopy (ARPES) is the most direct technique to access the electronic structure of materials. From comparison of experiments using soft X-ray and VUV radiation it has become clear that there is a close connection between 2DEGs found at interfaces and those found at surfaces, although the different environment will cause subtle differences in the electronic structure \cite[and references therein]{Cancellieri:2015,Plumb:2017,Frantzeskakis:2017}. In summary, the 2DEG consists of two circular mainly $3d_{xy}$-derived states with clear 2D characteristics and low effective mass ($m^*=0.65m_e$), and ellipsoidal $3d_{xz}$- and $3d_{yz}$-derived states with much higher effective mass ($m^*=15m_e$) and 3D-like dispersion (Fig.~\ref{fig.1}a). The formation of the 2DEG is most likely related to a cobination of oxygen vacancies, structural distortions, and a confining potential at the surface, while its population in addition strongly dependents on light-induced effects \cite{Santander:2011,Meevasana:2011,Plumb:2014,Walker:2015}. More recently, ARPES measurements on STO film grown by molecular beam epitaxy (MBE) suggested that the formation of the 2DEG correlates with the SrO surface termination \cite{Rebec:2019}, while previous ab-initio calculations also indicated that an SrO-terminated surface favors charge accumulation \cite{Delugas:2015}.

Another promising aspect of the systems described above has been the discovery of a gate-tuneable Rashba-type effect with a Rashba parameter $\alpha$~=~3.4$\times 10^{-12}$~eVm for the 2DEG at LaAlO$_3$/STO  (LAO/STO) interface \cite{Caviglia:2010}, while LaTiO$_3$/STO interfaces have shown a larger splitting of 1.8$\times 10^{-11}$~eVm \cite{Veit:2018}. For the 2DEG at the SrTiO$_3$(001) surface an even larger $\alpha$~=~5$\times 10^{-11}$~eVm was found by spin- and angle-resolved photoemission spectroscopy (SARPES) \cite{Santander:2014}. The latter shows a helical Rashba-like spin texture in addition to a Zeeman-like gap around the surface Brillouin zone (SBZ) centre. The exact nature of this observation and the relative contributions of magnetic order and spin-orbit interaction are still under debate \cite{Gorkov:2015,Garcia:2016}. A further SARPES measurement under different conditions failed to reproduce the finding \cite{McKeown:2016}, most likely due to the chosen experimental parameters (see Appendix, Sec.\ref{SOM:sarpes} for details).

The proposed presence of both Rashba and Zeeman interactions leads to a spin gap $\Delta$ at the Brillouin zone centre of STO [Fig.~\ref{fig.1}(a)], promising a wide range of functionalities depending on where the chemical potential is placed. An interesting scenario occurs when the chemical potential is placed inside the Zeeman gap, in which case the system would show only one helical Fermi surface and resemble the situation required to form Majorana bound states \cite{Mourik:2012}. Especially given the presence of superconductivity in both bulk STO \cite{Schooley:1964} and STO interfaces \cite{Reyren:2007,Bert:2011}, a whole new realm of physics opens up based on the interplay of magnetism, spin-orbit interaction, and superconductivity in a single material.

As indicated above, the functionality mostly relies on the possibility to shift the chemical potential while not altering other properties. In semiconductors and oxides a common approach is to use a gate voltage, but for the LAO/STO interface this is known to also change the magnitude of the Rashba-type splitting \cite{Caviglia:2010}. Furthermore, it is unclear whether such an approach causes any shift in the surface 2DEG of STO(001). Remarkably, apart from the 2DEG generated by Al deposition \cite{Roedel:2016b}, all published ARPES studies on the surface 2DEG show an almost identical band filling irrespective of the bulk doping level and whether the sample is prepared by cleaving \cite{Santander:2011,Meevasana:2011} or \textit{in-situ} annealing \cite{Plumb:2014}, and of the amount of oxygen vacancies  in the bulk or a surface. Only at the initial moment of irradiation by photons \cite{Meevasana:2011}, by surface contamination \cite{Walker:2015}, or by significant structural alterations to the surface to obtain TiO$_2$ enriched SrTiO$_3$ \cite{Wang:2016}, the band filling can be altered for an instance. In these cases the carrier density is determined \textit{a posteriori} but can not be controlled in a stable and reproducible manner. This universal filling indicates that the origin of the 2DEG lies beyond a simple band bending picture, in which other ingredients such as structural distortions and surface termination may also play a role.

In this work we follow a different and stable approach to tailor the chemical potential, namely the homoepitaxial growth by pulsed laser deposition (PLD) of thin SrTiO$_3$ films on nominally TiO$_2$-terminated Nb-doped SrTiO$_3$(001) substrates. The use of this rather standard growth technique makes our results directly applicable to the use of STO films in epitaxial systems. Our main finding is that for films from 3 to at least 20 unit cells (u.c.) on 0.5 wt.\% Nb-doped substrates the band filling is such that the Fermi level is exactly in the Zeeman gap at the SBZ centre, thus resulting in a single spin-polarised 2D Fermi surface (Fig.~\ref{fig.1}b). Further, we show that the position of the Fermi level varies with the amount of Nb dopants in the substrate. The similarities of the 2DEG found in films with different thicknesses indicate that the difference between single crystals and the films is not due to finite size effects. 

\section{Results}

\subsection{Films on highly-doped substrate}

ARPES experiments were performed at the surface and interface spectroscopy beamline of the Swiss Light Source with the sample held at a temperature below 20~K. At initial exposure of the sample to the beam, we do not observe any intensity at the Fermi level. During the exposure, we note that parabolic states start to develop with a continuous increase in signal intensity. In contrast to other systems \cite{Meevasana:2011,Walker:2015,Rebec:2019} we observe no shift of the band position but only an increase in intensity. All the following data were acquired after saturation of the signal. 

 \begin{figure}[htb]
	\includegraphics[width=0.8\columnwidth]{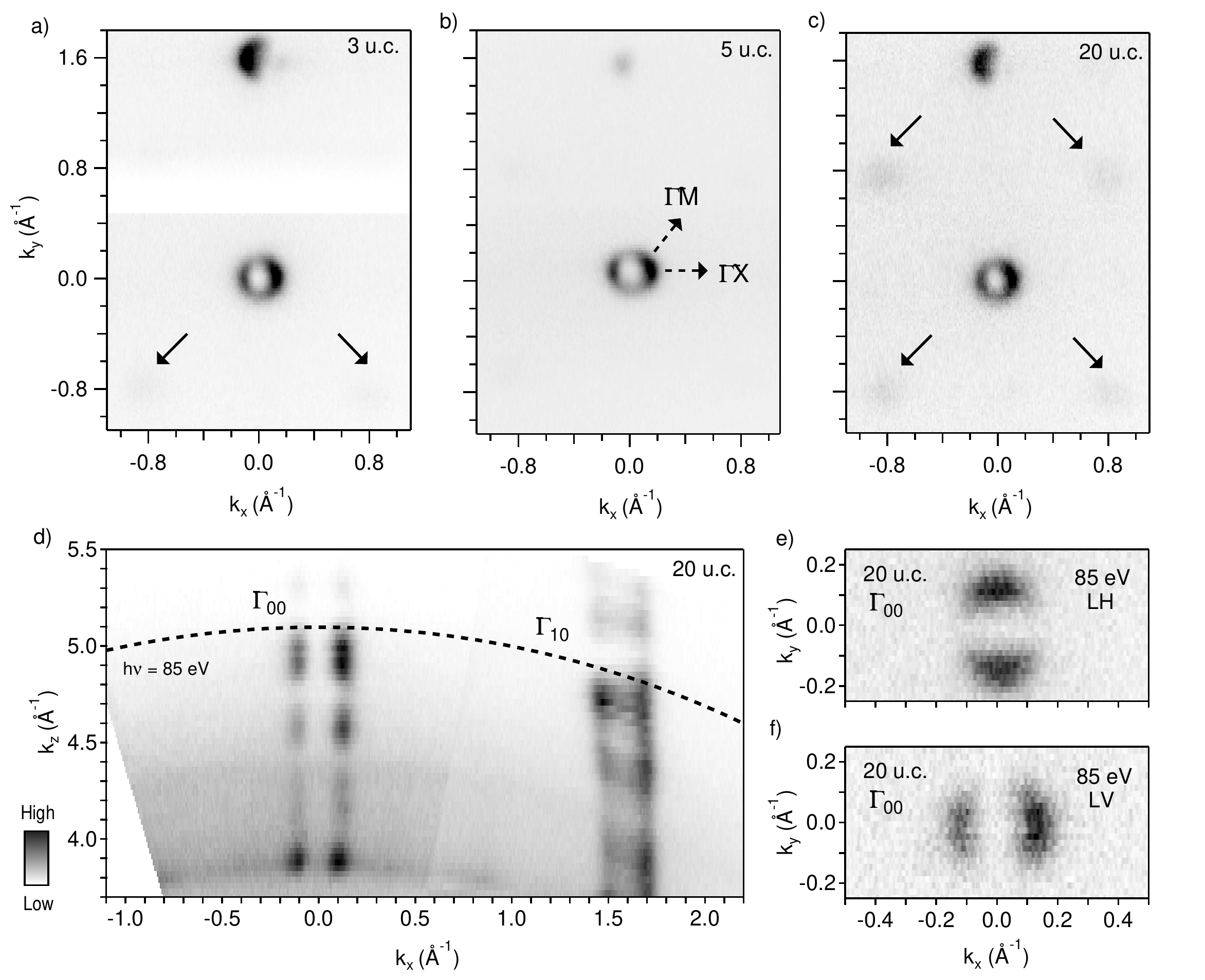}
	\caption{3D Fermi surface mapping of state at SrTiO$_3$ film surfaces. (a-c) Fermi surface of 3, 5, and 20 u.c. thick films grown on the 0.5 wt\% Nb-doped substrate, measured at $h\nu$~=~85~eV with circularly polarised light (C$^+$) covering $\overline{\Gamma_{00}}$ and $\overline{\Gamma_{01}}$ as well as reconstructed $\Gamma$-points. The lack of signal in (a) at $k_y = 0.5$\AA$^{-1}$ is due to a synchrotron beam loss. (d) Dispersion with $k_z$ for the 20 u.c. sample along $\overline{\Gamma \mathrm{X}}$. (e,f) Zoom of FS with linear-horizontal (LH) and linear-vertical (LV) polarizations.}
	\label{fig.hv}
\end{figure}

The ARPES data for the 3, 5, and 20 u.c. STO films on 0.5 wt\% Nb-doped STO substrate shows marked differences to the data typically obtained for STO single crystals \cite{Santander:2011,Meevasana:2011,Plumb:2014}. In Fig.~\ref{fig.hv}(a-c), circular Fermi surfaces are observed around the $\Gamma$ points and there is no signature of the ellipsoidal $d_{xz}$- and $d_{yz}$-derived states for any of the three film thicknesses. The absence of the heavy bands is verified with the 20 u.c. film in the photon energy scan (\textit{i.e.} dispersion along k$_z$, calculated using an inner potential V$_0$~=~14.5~eV~\cite{Plumb:2014}), shown in Fig.~\ref{fig.hv}(d), which shows a pure 2D character. The observed intensity variations are due to the 3$p$--3$d$ resonance at around $h\nu$~=~45~eV (k$_z \approx$~3.8~\AA$^{-1}$) and Bloch spectral enhancement at bulk $\Gamma$ points\cite{Dil:2004}. However, a band folding, as sign of surface reconstructions,  is visible in all the Fermi surfaces with varying clarity. Further, that the 2DEG is $d_{xy}$-derived can be seen by dependence of the Fermi surface with light polarisation (linear vertical and horizontal, LV and LH), which in our experimental geometry is a signature of a band with $d_{xy}$ character [Fig.~\ref{fig.hv}(e,f)].

Signatures of a purely 2D state are also seen in the 5 u.c. film (see  the Appendix, Sec.\ref{SOM:dim} for details). For all studied films, there is no sign of the heavy bands at 85~eV , which is the typical energy at which these states are expected to appear in STO crystals \cite{Plumb:2014}. Given the similarities, we assume all the films to host a purely $d_{xy}$-derived 2D state. Finally, as evident also in their RHEED pattern (details in the Appendix, Sec \ref{SOM:growth}), we observe a $\sqrt{2}\times\sqrt{2}\mathrm{R}45$\degree~surface reconstruction, indicated by the solid arrows in Fig.~\ref{fig.hv}(c). This type of surface reconstruction in STO has been assigned to SrO at the surface \cite{Ogawa:2017}. Despite their visibility in the ARPES spectra, the bands around the reconstructed $\overline{\Gamma}$ points are relatively featureless (see the Appendix, Sec. \ref{SOM:dim}). This suggests that the reconstruction is not long range ordered, and thus was not considered in the following analysis.

\begin{figure}[htb]
	\includegraphics[width=0.8\columnwidth]{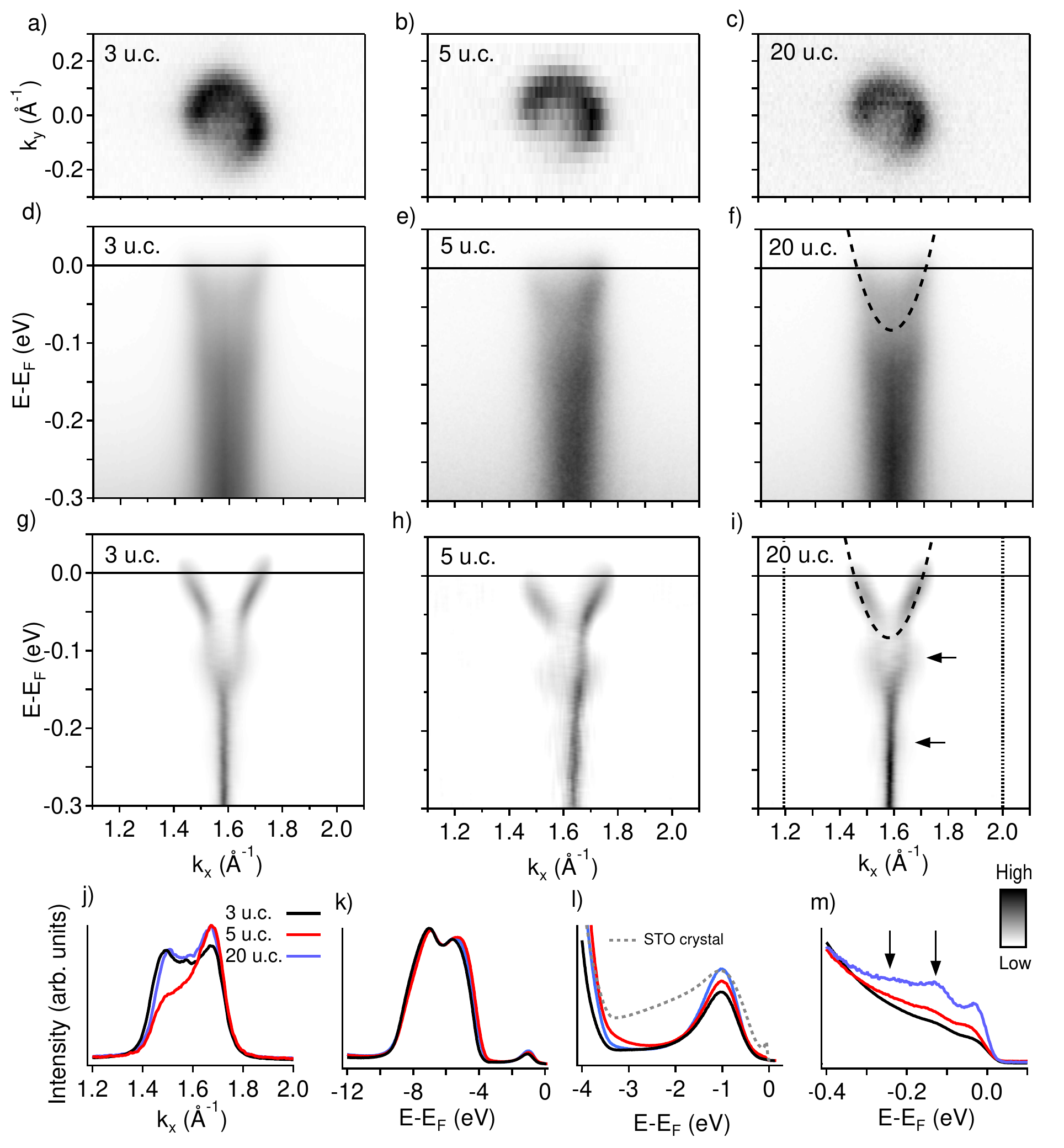}
	\caption{The 2DEG on 10 u.c. film grown on highly-doped substrate. (a) Fermi surface, (b) band structure and (c) 2D curvature for 3 u.c. STO on the 0.5 wt\% Nb-doped substrate at $\overline{\Gamma_{01}}$ with $h\nu$=85~eV (C$^+$). (b,e,h) and (c,f,i) same as (a,d,g) but for 5 u.c. and 20 u.c. respectively. A free electron like parabola is indicated in (f,i). (j) MDCs of the three films at the Fermi energy from (d,e,f). $k_x$ integrated EDCs of the valence band (k), the in-gap state (l), and close to the Fermi level (m) with indicated polaron replicas (black arrows). In (l) the spectrum of a STO single crystal is shown for comparison.}
	\label{fig.arpes}
\end{figure}

A close comparison of the band structures of the 3, 5, and 20 u.c. STO films on 0.5 wt\% Nb-doped STO substrate shows very similar Fermi surfaces [Figs.~\ref{fig.arpes}(a-c)] and band dispersions [Fig.~\ref{fig.arpes}(d-f)]. There is no noticeable change in Fermi wave vector $k_F$ [Fig.~\ref{fig.arpes}(j)], and the formation of the valence band [Figs.~\ref{fig.arpes}(k)] and the in-gap state [Figs.~\ref{fig.arpes}(l)] are also similar for the three films. In the 2D curvature [Figs.~\ref{fig.arpes}(g-i)] \cite{Peng:2011} and in the $k_x$ integrated EDCs [Figs.~\ref{fig.arpes}(m)], polaron replicas of the $d_{xy}$ band with an energy separation of $\approx$~100~meV are visible. These are large polarons formed in the photoemission process commonly observed in titanates \cite{Moser:2013, Wang:2016}. It is also worth noting the intense incoherent spectral weight at the center of the SBZ that appears below 150~meV, also often observed for the 2DEG on STO crystals \cite{Santander:2011,Meevasana:2011,Plumb:2014}. 

The fact that no differences are observed as a function of film thicknesses rules out the influence of finite size effects and constrains the spatial extension of this 2DEG from the top TiO$_2$ layer to 2 u.c. or less into the film and further hints at an origin beyond a simple band bending model \cite{Santander:2011,King:2014}. The 2D curvature data shows the shape of the $d_{xy}$ band that follows a free-electron-like dispersion. The parabola plotted in Figs.~\ref{fig.arpes}f) and (i), with a band bottom of 80~meV and an effective mass of m$^*$=0.74$m_e$ matches well the observed dispersion of the three samples. Furthermore, the small increase in effective mass (m$^*$=0.65$m_e$ for the 2DEG on STO single crystals \cite{Plumb:2014}) points to an altered bond angle, likely due to surface relaxation. Due to the absence of the heavy $d_{xz}$/$d_{yz}$ bands, it cannot be confirmed whether the splitting between these and the $d_{xy}$ band has changed. 

In addition to the reduced band filling, the in-gap states observed in our thin films are different from the ones observed in the bulk counterpart (Fig.~\ref{fig.arpes}m). For single crystal STO (dashed line) two in-gap states are observed whereas for the STO films the state at 2.5~eV binding energy has disappeared. Both in-gap states are known to originate from defects \cite{Kim:2009,Chambers:2018}. The different in-gap states suggests a different defect structure of our PLD-grown STO films when compared to STO crystals.

\begin{figure}[htb]
	\includegraphics[width=0.8\columnwidth]{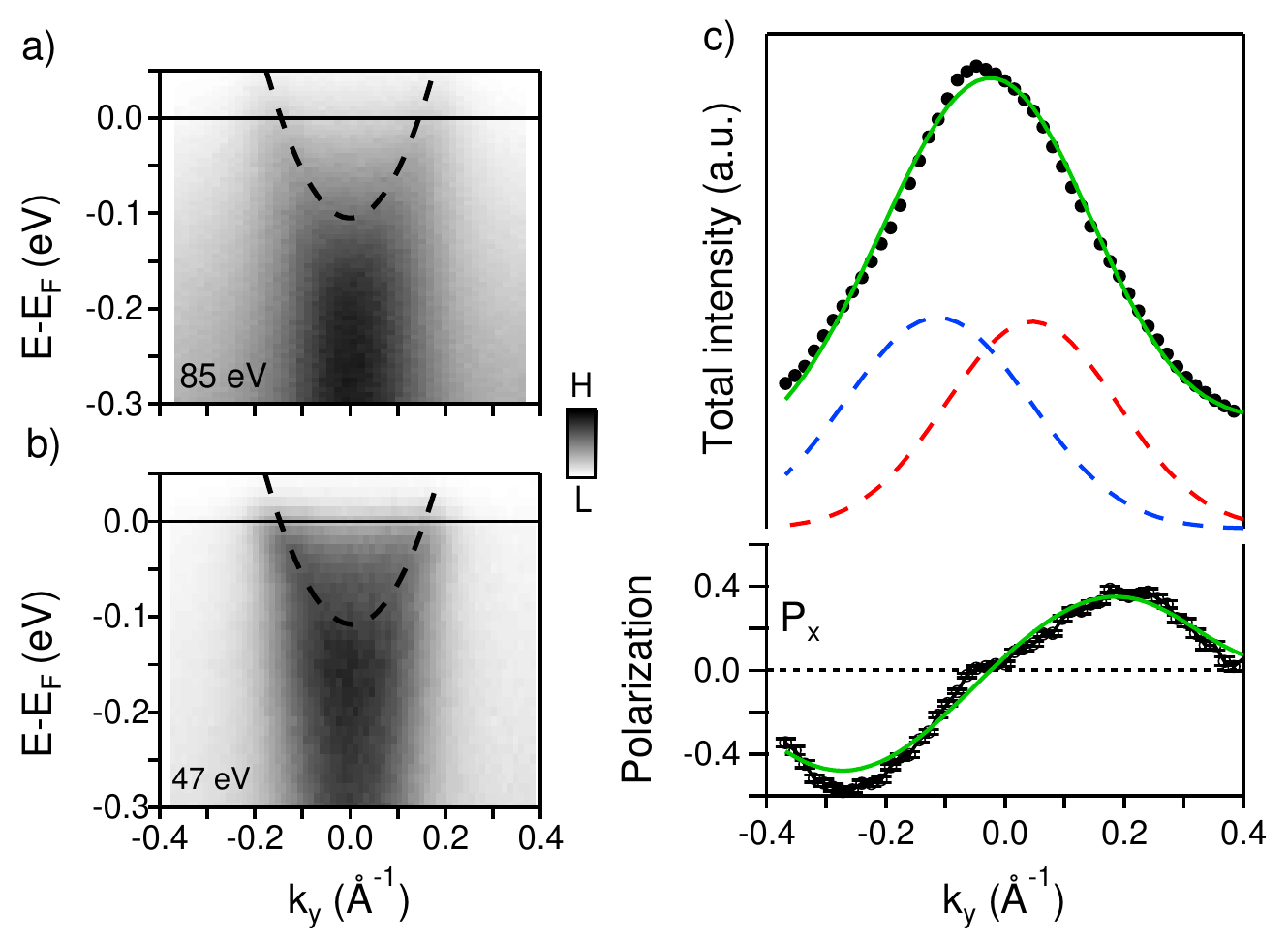}
	\caption{Spin-polarisation of single band at Fermi level. (a) Electron dispersion of 10 u.c. STO film grown on 0.5 wt\% Nb:STO with $h\nu=85 eV$ and (b) $h\nu=47 eV$. (c) Spin-resolved MDCs at the Fermi level and respective P$_{x}$ spin polarisation The solid green lines are results of the simultaneous fit of total intensity and spin-polarisation.}
	\label{fig.sarpes}
\end{figure}

Apart from the likely small change in bond angle, the most striking difference of the 2DEG found on our STO/Nb:STO films when compared to the universal 2DEG found in cleaved and annealed STO crystals \cite{Santander:2011,Meevasana:2011,Plumb:2014} is the large reduction of band filling from 230 to 80~meV. This corresponds to a downwards shift of the Fermi level, which now is crossed by a single band. (Fig.~\ref{fig.arpes}). Considering the spin texture measured for the 2DEG on the surface of bulk STO \cite{Santander:2014}, applying a rigid upwards energy shift would lead to a single spin-polarized Fermi surface. However, given the general differences of thin films, a change in spin splitting cannot be excluded and requires experimental verification. In order to access this often elusive degree of freedom, we employed spin-resolved ARPES to study a 10 u.c. STO/0.5~wt\% Nb:STO film, well within the previously studied 3--20 u.c. range.

Figure \ref{fig.sarpes} shows the spin-integrated band dispersion measured at the COPHEE end-station \cite{Hoesch:2002} for a 10 u.c. STO/0.5~wt\% Nb:STO film, along the $\overline{\Gamma \mathrm{Y}}$ direction, with C$^+$ photons of 85~eV and 47~eV [Fig.~\ref{fig.sarpes}(a,b)]. Although the observed bottom of the d$_{xy}$ band in these measurements are 20 meV lower than in Fig.~\ref{fig.arpes}, still only one band is visible. The dashed blue line represents the band dispersion in Fig.~\ref{fig.arpes}, with the 20 meV shift taken into account. The spin-resolved MDC measured with $h\nu=$~85~eV photons at the Fermi level is shown in Fig.~\ref{fig.sarpes}(c). The main spin polarisation signal points along the $x$-direction, while the measured out-of-plane spin polarisation is most likely due to spin interference during the photoemission process \cite{Dil:2019} and $|P_y|\le0.04$ (see the Appendix, Sec. \ref{SOM:comp_sarpes} for details).

A well-established routine \cite{Meier:2009NJP} was used to simultaneously fit the total intensities and the spin polarisation data. The fitted total intensity and polarisation along the sample $x$ direction (P$_x$) are represented by green solid lines in Fig.~\ref{fig.sarpes}(b), while the red and blue dashed lines represent the individual peaks of the fit. The spectrum around $\overline{\Gamma_{10}}$ originates from a single band whose polarisation is perpendicular to the crystal momentum and reverses sign at the SBZ centre, consistent with a helical spin texture as also observed for the 2DEG on STO crystals \cite{Santander:2014}. Hence, the Fermi level for the films grown on 0.5 wt\% Nb-doped STO(001) substrate [Fig.~\ref{fig.arpes}] lies inside the Zeeman gap and the electronic structure thus show a single spin-polarised Fermi surface.

\subsection{Films on low-doped substrate}

\begin{figure}[htb]
	\includegraphics[width=0.9\columnwidth]{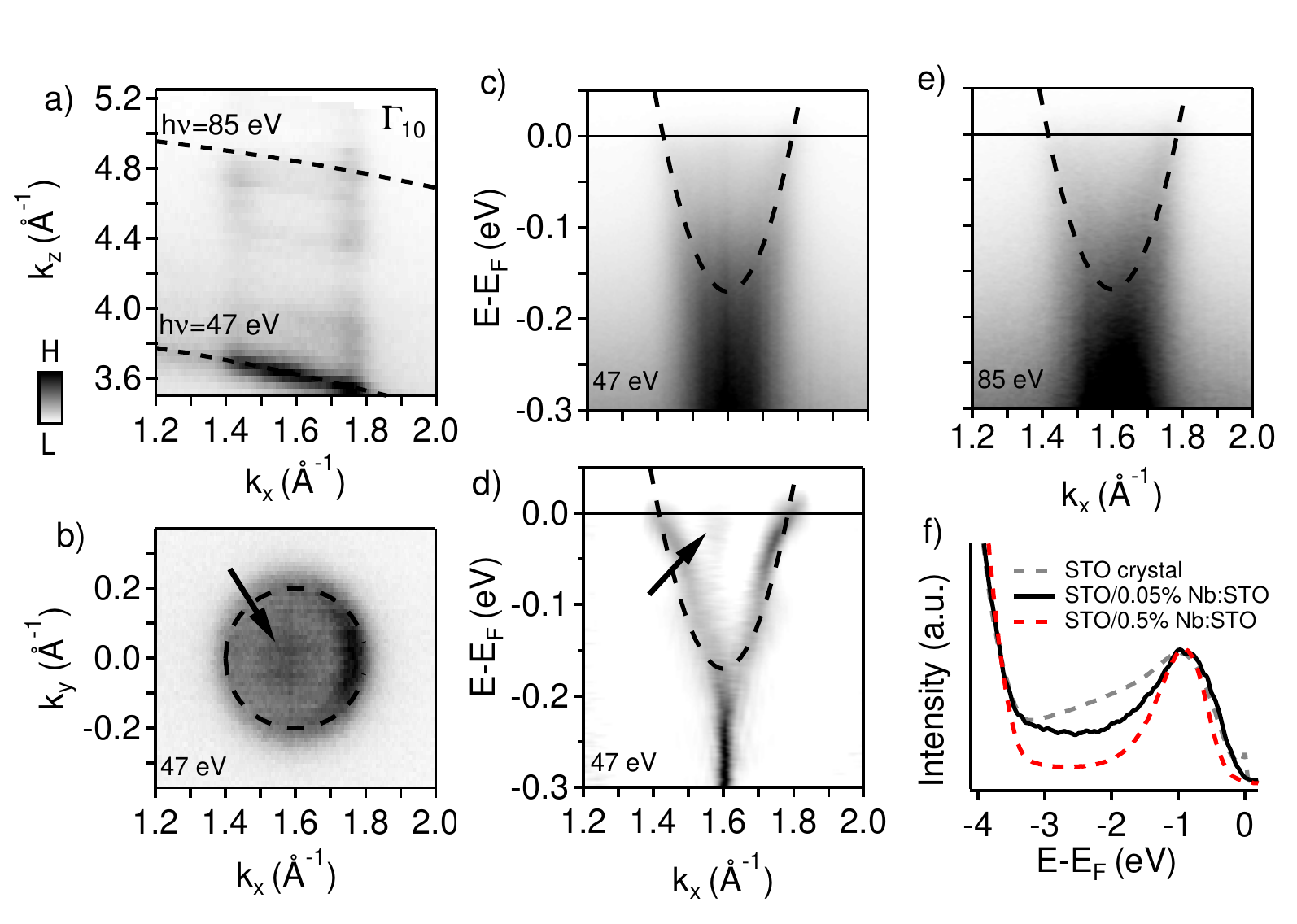}
	\caption{The 2DEG on film grown on low-doped substrate. (a) Fermi surface in the k$_x$-k$_z$ plane and (b) k$_x$-k$_y$ plane of 10 u.c. STO film grown on 0.05 wt\% Nb:STO. (c) Band dispersion along $\overline{\Gamma \mathrm{X}}$ measured with $h\nu = 47 eV$ and the respective second derivative analysis (d). (e) Band dispersion along $\overline{\Gamma \mathrm{X}}$ measured with $h\nu = 85 eV$. (f) EDC of the in-gap states, where the spectrum from a STO crystal is shown for comparison.}
	\label{fig.lowdoped}
\end{figure}

In order to further explore the 2DEG developed on PLD-grown STO/Nb:STO(001) films, 
we grew 10 u.c. films on 0.05 wt.\% substrates, i.e., where the amount of Nb dopants is changed by one order of magnitude. Fig.~\ref{fig.lowdoped}(a) shows the k$_z$ dispersion around $\overline{\Gamma_{10}}$ for this film. As for the films on the low-doped substrate, here the heavy bands are not observed either. Fig.~\ref{fig.lowdoped}(b) shows a Fermi surface obtained with $h\nu$ = 47 eV C+, where a circular outer band with k$_F$ = 0.18~\AA$^{-1}$ is clearly evident, along with an increase in intensity towards the center. The band dispersion measured at k$_y$~=~0~\AA$^{-1}$ [Fig.~\ref{fig.lowdoped}(c)] and the respective curvature analysis [Fig.~\ref{fig.lowdoped}(d)] \cite{Peng:2011} reveal a band filling of around 170 meV. Again relying on the 2DEG on STO crystals, a rigid band shift of the outer d$_{xy}$ band from the 230 meV to the 170 meV would result in an inner d$_{xy}$ band populated up to $\approx$ 40 meV,  with k$_F$ $\approx$ 0.05~\AA$^{-1}$. Indeed, spectral weight can be seen around (k$_x$, k$_y$) $\approx$ (1.55, 0) \AA$^{-1}$ in Fig.~\ref{fig.lowdoped}(b,c,d), although a clear band dispersion could not  be fully resolved. Despite this uncertainty, which inspires further investigation with other differently doped substrates, it is clear that this film also does not show the heavy bands and hosts distinct (similar) in-gap states as in crystals (films on highly-doped substrates) [Fig.~\ref{fig.lowdoped}(f)].

\subsection{X-ray photoelectron spectroscopy of core-levels}

\begin{figure}[htb]
	\includegraphics[width=1\columnwidth]{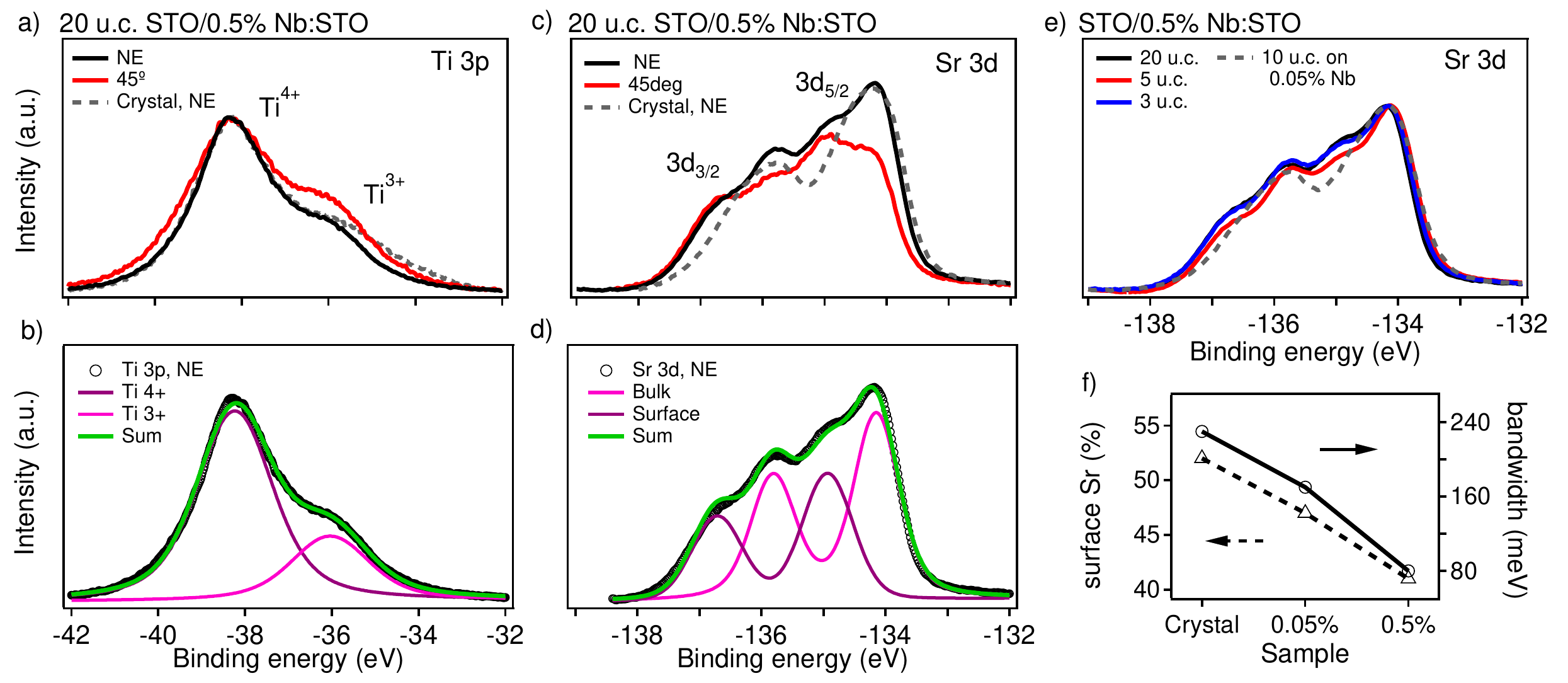}
	\caption{XPS analysis of PLD-grown STO films. (a) Ti~3p core-level measured at normal emission (NE) and 45\degree, as well as the spectrum of a STO crystal measured at NE,  (b) fit of the spectrum of the film at NE. (c,d) The equivalent dataset for Sr~3d core-level. (e) Comparison of the Sr~3d core-level of films of different thicknesses and substrates. (f) Correlation between band filling and the magnitude of the surface Sr contribution.}
	\label{fig.XPS}
\end{figure}

Our results raise the question why PLD-grown STO films on Nb:STO substrates show a smaller band filling. Further information can be gathered from a detailed analysis of Ti~3p and Sr~3d core-levels, shown in Fig.~\ref{fig.XPS}, measured with $h\nu$=~170~eV. In Fig.~\ref{fig.XPS}(a), the Ti 3p core-level of the 20 u.c. film grown on the 0.5\% Nb:STO substrate, measured at normal emission (NE), shows the typical Ti$^{4+}$ peak (which comprises the Ti~3p$_{3/2}$ and Ti~3p$_{1/2}$ contributions), followed by a Ti$^{3+}$ shoulder. Fig.~\ref{fig.XPS}(b) shows a fit of this data, evidencing the two components and allowing a quantitative analysis of the spectrum. The increase in Ti$^{4+}$/Ti$^{3+}$ ratio in the data measured with an emission angle of 45\degree~ shows that the Ti$^{3+}$ is mostly located at the surface of the sample, as previously observed \cite{Plumb:2014}. For comparison, in Fig.~\ref{fig.XPS}(a) we show the spectrum of a STO crystal, which is similar to the 20 u.c. film, apart from a wider Ti$^{3+}$ component. All the films studied in this work present Ti 3p core-levels with similar line shapes, and although the Ti$^{3+}$ content varies from sample to sample, it does not seem to show a systematic evolution with the Nb content or thickness (see details in the Appendix, Sec. \ref{SOM:xps}).

The Sr 3d spectrum of the 20 u.c. film  measured at NE [Fig.~\ref{fig.XPS}(c)] also shows a two-component shape, which is decomposed in Fig.~\ref{fig.XPS}(d) including the 3d$_{5/2}$ and 3d$_{3/2}$ components. The spectrum measured at 45$\degree$ emission confirms that the component at higher binding energy is mostly located at the surface of the sample, and is typically attributed to the formation of SrO$_x$ crystallites \cite{Hatch:2013}. Furthermore, an elemental analysis of the XPS spectra reveals that the Sr/Ti ratio in our films is between 1-1.20, while the crystals we studied show a ratio of 1.4. These values are consistent with what was reported recently for MBE films grown with different terminations \cite{Rebec:2019}.

Comparing the Sr 3d spectrum of the 20 u.c. film with the one of a STO crystal, also shown in Fig.~\ref{fig.XPS}(c), we observe not only an increase in intensity of the surface Sr contribution, but also an energy shift of about 200~meV towards higher binding energy. Additionally, the surface Sr signal appears to be almost independent of film thickness, as seen in Fig.~\ref{fig.XPS}(e), where the 3, 5, and 20 u.c. films show very similar spectra. In particular, the spectrum for 5 u.c. shows a slightly smaller surface Sr contribution, although the peak position matches well the ones of the 3 and 20 u.c. films. For further comparison, Fig.~\ref{fig.XPS}(e) also presents the Sr 3d spectrum of the 10 u.c. 0.05\% Nb:STO film, which differs from the films on the highly-doped substrate, and more closely resembles the one from STO crystals [Fig.~\ref{fig.XPS}(c)].

We now recall that the filling of the 3d$_{xy}$ band varies from crystals to films with either doping, although the amount of Ti$^{3+}$ is similar in all samples. This indicates that there is no direct correlation between the amount of Ti$^{3+}$ with the band filling of the 2DEG, likely due to the fact that the extra charges in STO not only form a 2DEG, but also remain localized around defects and show up as the in-gap states [Figs.~\ref{fig.arpes}(l)] \cite{Janotti:2014,Hao:2015}. In turn, the shape of the Sr 3d spectra and the amount of SrO$_x$ crystallites shows a trend when compared to the observed band filling, as seen in Fig.~\ref{fig.XPS}(f). Whether this observation fully explains the differences in band filling requires further investigation with microscopy and spectromiscroscopy techniques. \cite{Sokolovic:2019}

As mentioned previously, the observed differences regarding in-gap states suggest that defects present in our thin films are different than those in STO crystals. In fact, this may be another reason for the change in band filling from wafers to thin films, since a different defect structure leads to a different dielectric response of each system. This, in turn, is expected to alter the properties of the polaronic excitations and thus the band characteristics \cite{Stengel:2011}. Optical and spectroscopic measurements have shown that the dielectric response of SrTiO$_3$ changes from single crystals to thin films \cite{Sirenko:2000,Ostapchuk:2002}, which can ultimately impact the electronic confinement and the observed band filling. In this respect it is important to note that the MBE grown films on 0.05\% Nb-doped substrates \cite{Rebec:2019} show a band bottom at around 170 meV, which is comparable to our results on PLD films grown on similar substrates. The above hypothesises about the influence of SrO$_x$ at the surface and the dielectric response are not mutually exclusive, and may indeed be closely related.

\section{Discussion and conclusions}

As indicated above, we found that the band filling for the film grown on the 0.5\% Nb:STO substrate is such that the chemical potential lies inside the Zeeman gap, leading to a single spin-polarised band (Fig.~\ref{fig.1}b). In practice, this result hints that local gating of the substrate can be used to create wires at whose tips zero bias anomalies might be observable by tunnelling experiments. Alternatively the wires can also be written by illuminating with an intense light source, by local defect doping, or by writing with a conductive tip \cite{Cen:2008}. In combination with the superconducting properties of STO, this unifies all the ingredients for the formation of Majorana bound states in a single material without the need of external fields.

The indication that the doping level of the substrate can influence the band filling of the 2DEG - and not in a trivial manner, since the higher the doping level in the substrate, the smaller the band filling - inspires further systematic and detailed study. If shown to be true, this would allow to tune the 2DEG on STO in a very stable way, and open the way to pre-patterning the substrate, enabling the growth of regions with different band filling (and topology).

\section*{ACKNOWLEDGMENTS}

This work was financially supported by the Swiss National Science foundation (SNF) Project No. PP00P2\_144742 and No. PP00P2\_170591. Z.W. was supported by the National Natural Science Foundation of China No.11874367.

\appendix*
\section{{\label{SOM}Supporting information}}
\renewcommand{\thefigure}{A\arabic{figure}}
\setcounter{figure}{0}


\subsection{Growth of the films\label{SOM:growth}}

\begin{figure}[htb]
	\includegraphics[width=0.8\columnwidth]{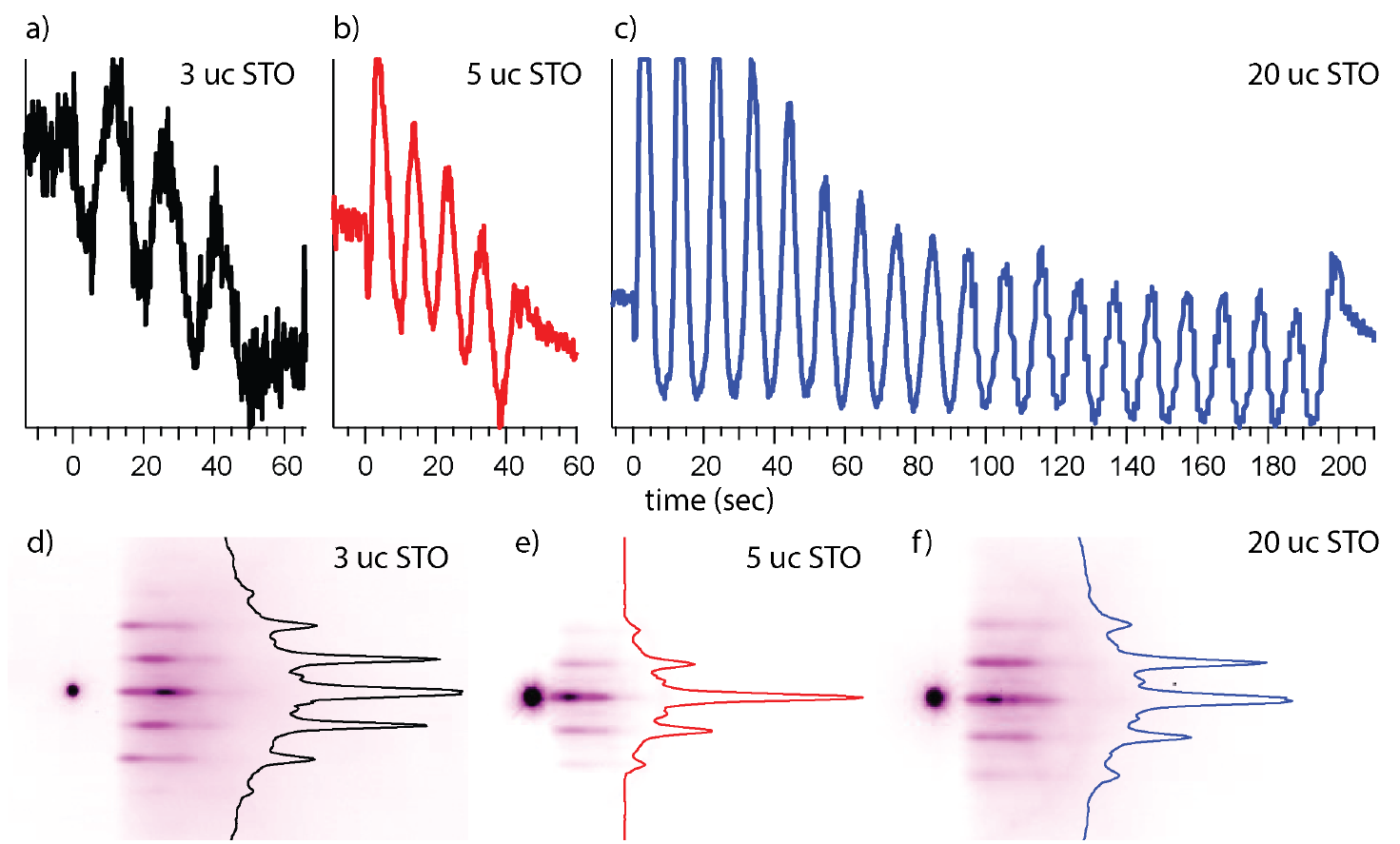}
	\caption{RHEED oscillations of the film growth of (a) 3 u.c., (b) 5 u.c. and (c) 20 u.c. STO on 0.5 wt\% Nb:STO(001) substrates. (d-f) RHEED patterns of the three films after the growth at the growth temperature, along with the intensity line profiles.}
	\label{fig.rheed}
\end{figure}

The SrTiO$_3$ thin films were grown by pulsed laser deposition (PLD) on commercially available, single TiO$_2$ terminated 0.05 and 0.5 wt\% Nb-doped STO(001) substrates (Twente Solid State Technology), using a crystalline STO disk as target. The reflection high-energy electron diffraction (RHEED) oscillations used to control the thicknesses of 3, 5, and 20 u.c. films on 0.5 wt\% Nb-doped STO(001) are displayed in Fig.~\ref{fig.rheed}(a-c). The RHEED patterns recorded after the growth show straight lines indicating a flat, 2D film surface. Between the main diffraction intensities, low intensity lines are visible [Fig. \ref{fig.rheed}(d-f)]. It is interesting to note that the RHEED intensity actually increases with the growth of the first overlayer of STO on the substrate, indicating the high quality of the films. 

\subsection{Details of (S)ARPES measurements\label{SOM:exp}}
 
 The prepared films were transferred \textit{in-situ} to the high-resolution ARPES endstation at the Surface and Interface Spectroscopy beamline of the Swiss Light Source at the Paul Scherrer Institut. X-ray photoemission (XPS) and ARPES spectra were measured with a Scienta R4000 analyzer with instrumental angle and energy resolution better than \ang{0.2} and 10~meV. In order to perform spin-resolved ARPES, new films were grown and transferred in a vacuum suitcase under ultra-high vacuum conditions to the COPHEE endstation \cite{Hoesch:2002}, which uses an Omicron EA~125 hemispherical energy analyser and two orthogonally assembled classic Mott detectors. The angle and energy resolutions of the SARPES measurements are better than \ang{1.5} and 70~meV. For all the measurements the samples were kept under pressures better than 5$\times$10$^{-10}$~mbar and at 20~K. No further sample treatment was required. These results were reproduced on different samples, in independent experiments.

\subsection{Dimensionality of the 2DEG\label{SOM:dim}}

\begin{figure}[htb]
	\includegraphics[width=1\columnwidth]{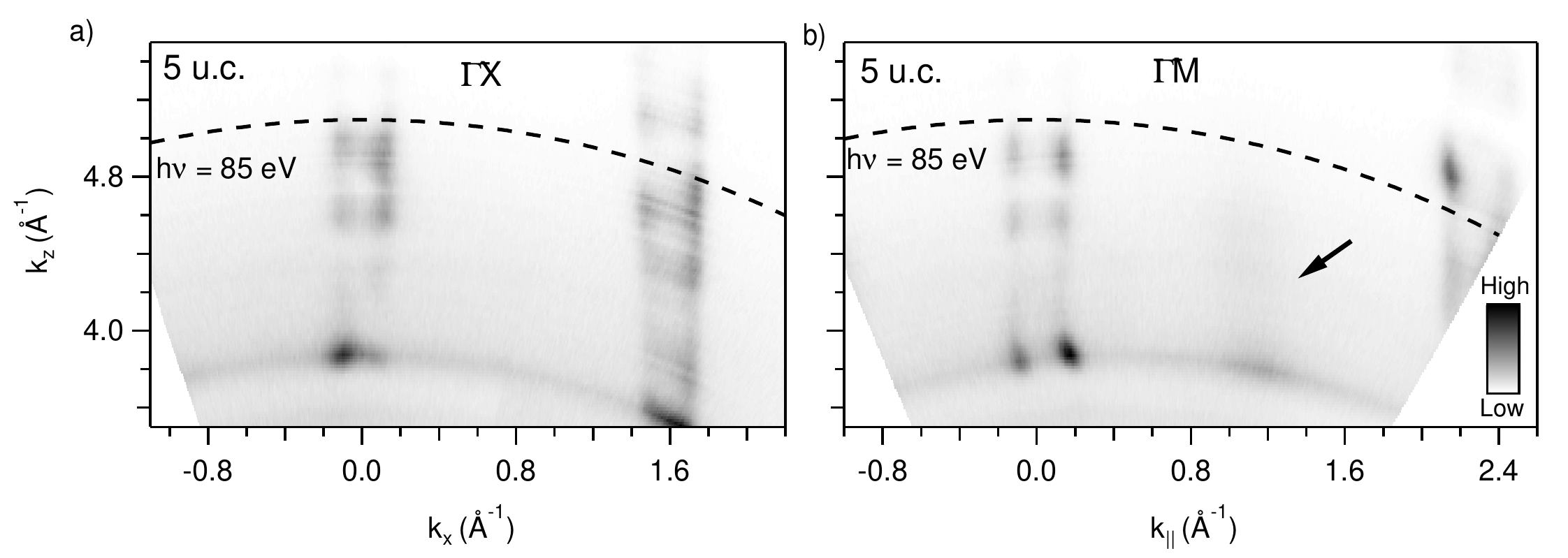}
	\caption{3D Fermi surface mapping of state at 5 u.c. SrTiO$_3$ film grown on the 0.5 wt\% Nb-doped substrate. Along (a) $\overline{\Gamma \mathrm{X}}$ and (b) $\overline{\Gamma \mathrm{X}}$, with circularly polarized light (C$^+$) covering $\overline{\Gamma_{00}}$ and $\overline{\Gamma_{01}}$. The arrow in (b) indicates the reconstructed $\overline{\Gamma}$ point.
	}
	\label{kz5uc}
\end{figure}

In the photon energy scans of the 20 u.c. film shown in Fig.~2 of the main text, the $d_{xy}$-derived state shows a pure 2D character. This is also concluded for the 3 and 5 u.c. films as noted below. Fig.~\ref{kz5uc} shows the a the Fermi surface in kx-ky plane for the 5 u.c. film along the symmetry direction GX and FM, which show a pure 2D character, with no hints of the heavy bands. Intensity variations are observed as in the 20 u.c. film presented in the main text. In particular for the GM direction, we observe spectral weight due to the $\sqrt{2}\times\sqrt{2}\mathrm{R}45$\degree~surface reconstruction, as pointed by the arrow in Fig.~Fig.~\ref{kz5uc}(b), as well as in the Fig.~2 of the main text. The bands around the reconstructed $\overline{\Gamma}$ points are relatively featureless, and apart from a resonant enhancement do not show the same structure as a function of photon energy as the main $\overline{\Gamma}$ points. It appears that the reconstruction is not long range ordered, and thus it was not considered in the analysis. Given the similarities of the data of the 3 u.c. film with the 5 and 20 u.c. ones, particularly the absence of the heavy bands in the spectra measured with $h\nu=85$~eV, we conclude that the 3 u.c. film also shows a purely 2D state.

\subsection{ARPES of films on different substrates\label{SOM:comp}}

The larger band filling of the 2DEG on the films grown on the 0.05 wt\%, compared to the ones and 0.5 wt\% Nb-doped STO(001) substrate, is presented in Fig.~5 of the main text. Here we compare the MDCs and EDCs obtained from these two samples, to show that the data indeed suggests the presence of 2 bands on the film grown on 0.05 the wt\% substrate, in contrast to the 1 band present in the film on 0.5 wt\% one.

\begin{figure}[htb]
	\includegraphics[width=0.6\columnwidth]{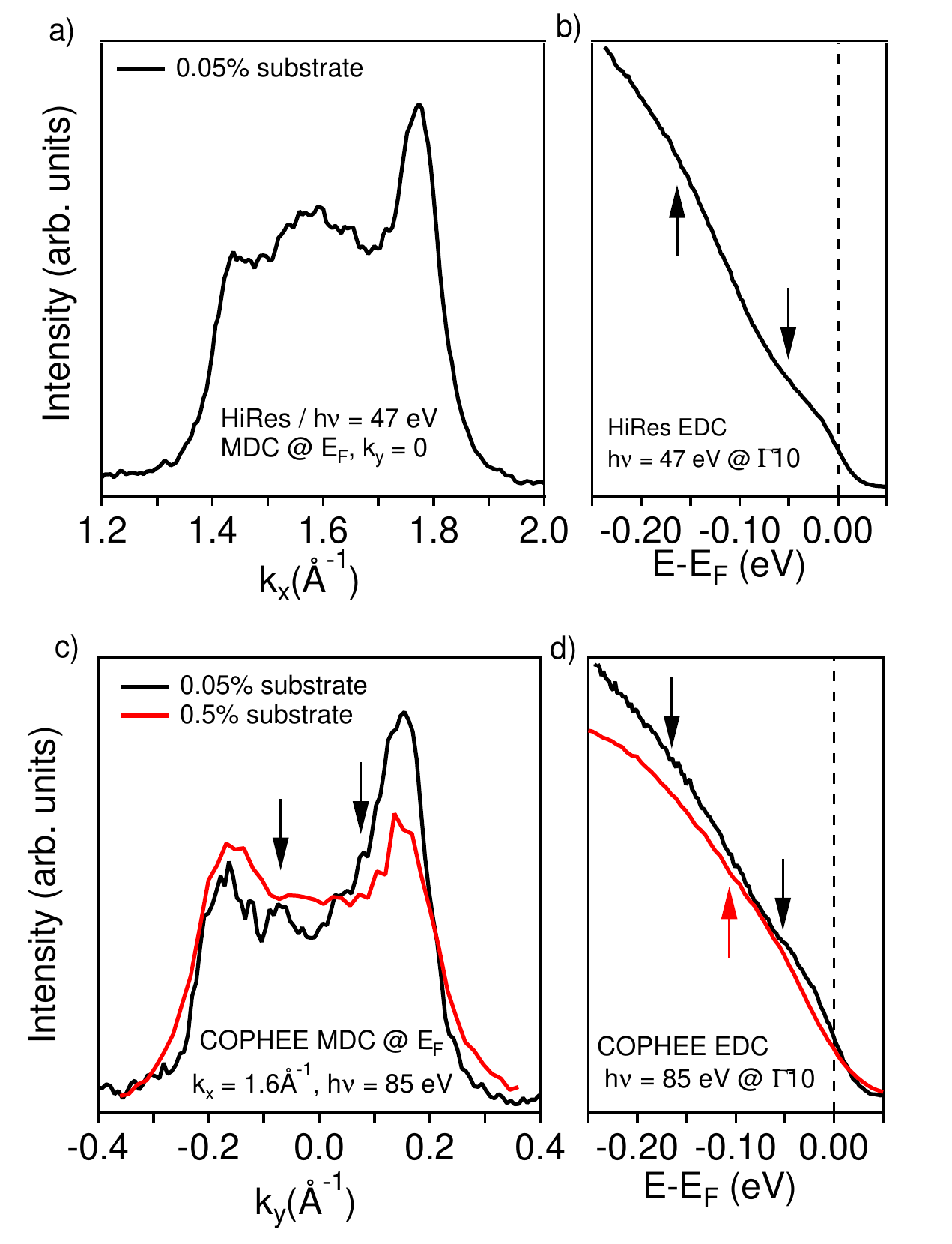}
	\caption{Extracted (a) MDC and (b) EDC for the 0.05\% sample measured at the high-resolution end station at 47 eV, C+. (c) and (d) shows MDCs and EDCs for the  0.5\% and 0.05\% Nb-doped samples measured at COPHEE with 85 eV photons, C+. 
}
	\label{MDCEDC}
\end{figure}

We begin by presenting the MDC and EDC from the data of Fig.~5 of the main text, \textit{i.e.} the 0.05\% sample measured at the HiRes end station with 47 eV photons. The MDC shows peaks at 1.43 and 1.77 \AA$^{-1}$, corresponding to the outer band, and a broad structure from 1.52 to 1.66 \AA$^{-1}$ [Fig.~\ref{MDCEDC}(a)]. In turn, the EDC (Figs~\ref{MDCEDC}(b)) shows bumps corresponding to the bottoms of the bands, at 50 and 170 meV. Note that EDCs at k$_y$=0 typically give only a low peak to background ratio in this system.

Now we discuss the EDCs and MDCs from the measurements with 85 eV photons performed at the COPHEE endstation. The MDCs at E$_F$ are shown in  Fig~\ref{MDCEDC}(c). The data for the 0.05\% sample shows an inner structures at k$_x$ = -0.07 \AA$^{-1}$ and k$_x$ = 0.07 \AA$^{-1}$ (marked with black arrows), which are absent in the spectrum of the 0.5\% sample. As for the EDCs in  Figs~\ref{MDCEDC}(d), despite the intense background, the data for the 0.5\% sample shows one structure between E$_F$ and 100 meV, while the 0.05\% sample shows structures at 50 and 170 meV, which correspond to the bottom of the bands observed in the angle-resolved spectra.

The differences are difficult to observe in the spin-integrated data obtained at the COPHEE end station. This is primarily because those measurements are performed using an old-fashioned channeltron detector with rather limited energy and angular resolution. The expected change in k$_F$ for a rigid band shift of 90 meV is around 0.05 \AA$^{-1}$, below its resolution limit (but above the one from HiRes endstation).

\subsection{SARPES of films on different substrates\label{SOM:comp_sarpes}}

\begin{figure}[htb]
	\includegraphics[width=0.6\columnwidth]{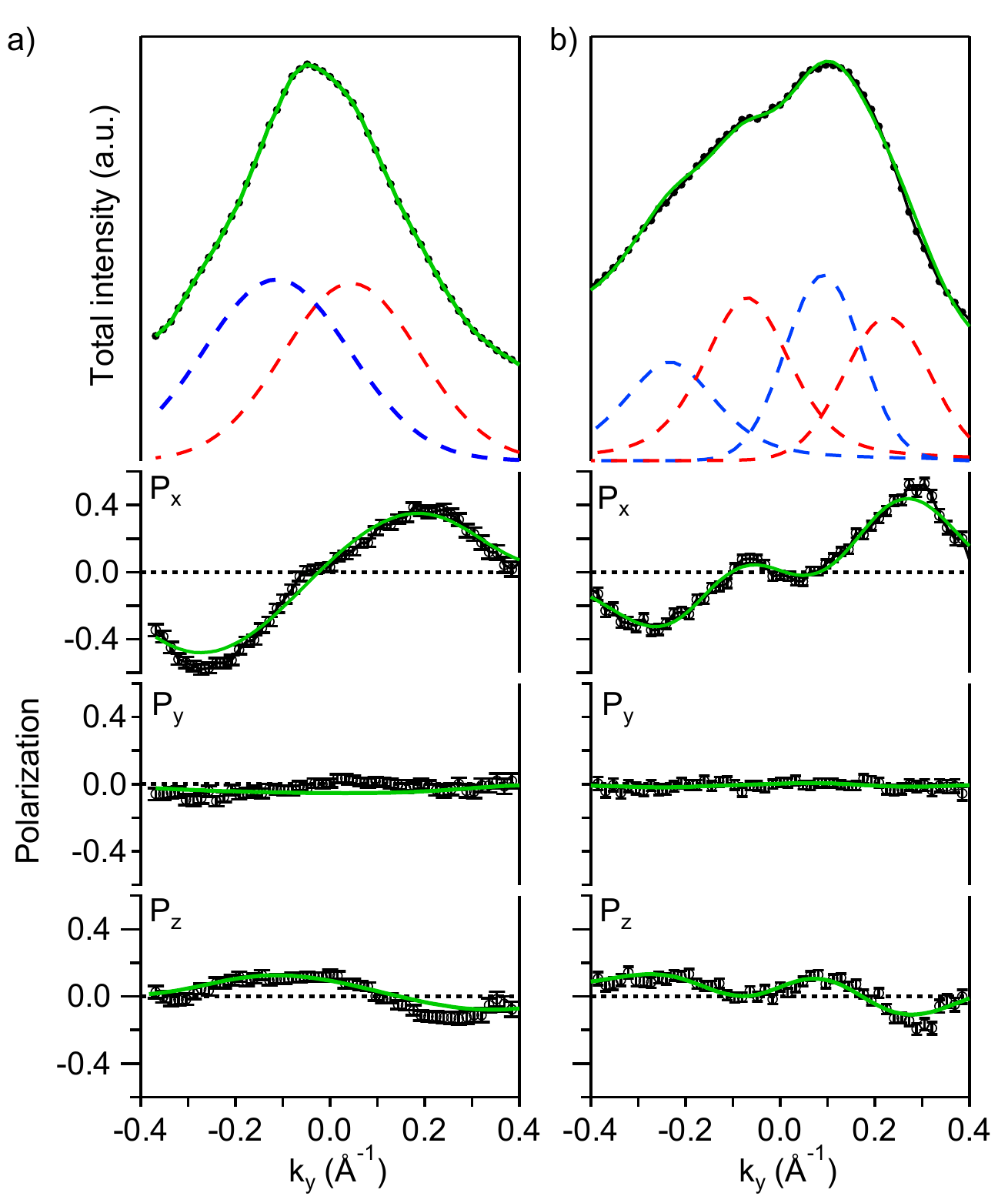}
	\caption{Spin-polarized MDCs at the Fermi level of two 10 u.c. films on 0.5 (a) and 0.05 (b) wt\% Nb-doped substrates. Each panel presents the total intensity split into the three spatial components P$_x$, P$_y$ and P$_x$ (markers), as well as the spin-resolved MDC (dashed lines). The green lines are the result of the simultaneous fit of the total intensity and P$_{x,y,z}$}
	\label{fig.sarpes_highdoped}
\end{figure}

We now proceed with the comparison of the obtained spin-resolved data. Fig.~\ref{fig.sarpes_highdoped}(a) shows the spin-resolved MDC of the 10 u.c. STO/0.5~wt\% Nb:STO film, measured at the Fermi level with C$^+$ photons of 85~eV, along the $\overline{\Gamma \mathrm{Y}}$ direction, the same as in Fig. 4(c) of the main text. As mentioned, the main spin polarization signal points along the $x$-direction, while $|P_y|\le0.04$ and $|P_z|\le0.2$. The measured out-of-plane spin polarization is most likely due to spin interference during the photoemission process \cite{Dil:2019}. The simultaneous fit \cite{Meier:2009NJP} of the total intensities and the spin polarizations along the sample x-, y-, and z-axis are denoted by the green lines. The fit resulted in a pair of peaks (one band), whose spin-polarization is consistent with a Rashba-type effect. Fig.~\ref{fig.sarpes_highdoped}(b) shows the equivalent data obtained for a 10 u.c. STO/0.5~wt\% Nb:STO film, grown under the same conditions as the film in Fig.~5 of the main text. In this case, the data could only be fitted with four of peaks (two bands), with oppositely winding spin textures, also consistent with a Rashba-type spin texture. This observation is consistent with the possibility of the second branch of the d$_{xy}$ band being populated, as discussed in the main text.

\begin{figure}[htb]
	\includegraphics[width=1\columnwidth]{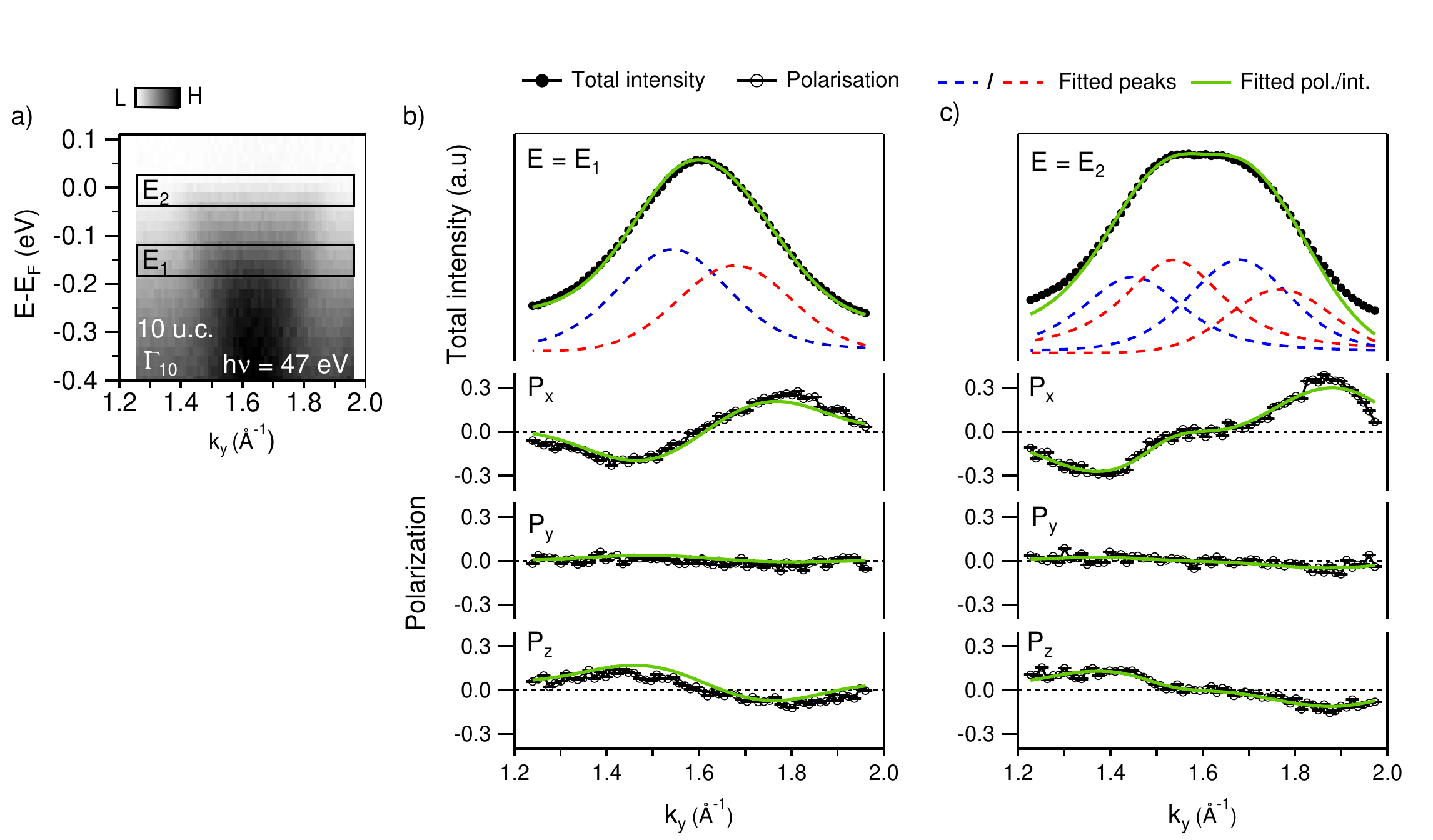}
	\caption{Spin-polarization of the 2DEG on a 10 u.c. film on 0.05 wt\% Nb-doped STO(001). (a) Spin-integrated band map measured with $h\nu$~=~47~eV, LV-polarized light, and the definition of the E$_{1,2}$ energy windows. Total intensity and related P$_{x,y,z}$ spin polarizations, along with their respective fits for (b) E~=~E$_1$ and (c) E~=~E$_2$.}
	\label{fig.sarpes_supp}
\end{figure}

We further explored this film with with $h\nu$~=~47~eV, vertically polarized light, measuring spin-resolved MDCs at two different binding energies E$_{1,2}$, defined in Fig.~\ref{fig.sarpes_supp}(a), and shown in Fig.~\ref{fig.sarpes_supp}(b,c). Again, for both E$_{1,2}$ the main spin polarization signal is along the $x$-direction, and the same origin for the polarization signals in y and z directions apply to this film. For E=E$_1$, the fit resulted in two peaks with opposite spin polarization, while the spectra measured around $E=E_2$ can only be fitted assuming four peaks with alternating spin polarization.

\begin{figure}[htb]
	\includegraphics[width=0.7\columnwidth]{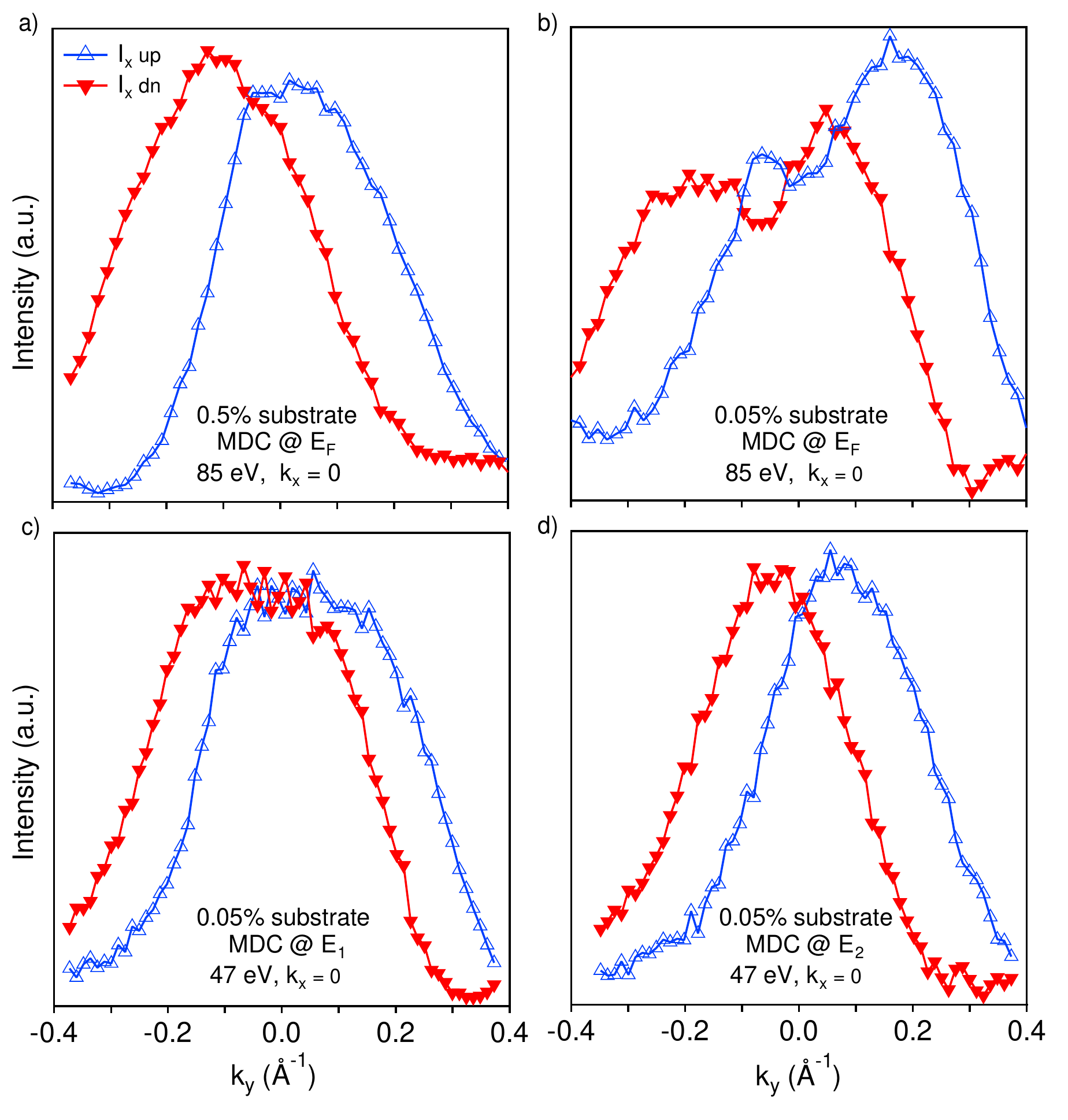}
	\caption{Spin-polarized MDCs projected on the x-axis, measured at COPHEE for (a) the 0.5\% Nb-doped sample, with 85 eV, C+ light, at E$_F$; (b) for the 0.05\% Nb-doped sample, with 85 eV, C+ light, at E$_F$; (c) for the 0.5\% Nb-doped sample, with 47 eV, LV light, at E$_2$;  (d) for the 0.5\% Nb-doped sample, with 47 eV, LV light, at E$_1$.}
	\label{fig.sarpes_supp2}
\end{figure}

For completeness and easy viewing we show in Fig.~\ref{fig.sarpes_supp2}(a,b) the spin up $I^\uparrow$ and down $I^\downarrow$ spectra projected on the $x$-axis for the data shown in Fig.~3 of the main text and Fig.~\ref{fig.sarpes_supp}. In  Fig.~\ref{fig.sarpes_supp2}(c,d) we show the equivalent data for the spectra presented in Fig.~\ref{fig.sarpes_supp}. The spectra are obtained by:
$$I_x^{\uparrow,\downarrow}=\frac{1}{2}\left(1\pm P_x\right)I_{tot}$$
Here $I_{tot}$ is the measured total intensity and $P_x$ the measured spin polarization along the sample $x$-axis. No further data treatment is applied and the number of peaks is easily resolved also from this graph. However, because of the projection on only a single axis, the splitting of the states and their relative intensity can be different compared to the analysis taking all three spatial components into account.

From the fact that the spin-resolved MDCs at E$_F$ of the films on 0.5\% and 0.05\% substrates [Fig.~\ref{fig.sarpes_highdoped}] look similar to when we change the binding energy, going from 1 to 2 bands [Fig.~\ref{fig.sarpes_supp}] (as well as to STO crystals \cite{Santander:2014}), and because we see the change in band filling in the high-resolution data, we infer that the the film on 0.05\%Nb:STO substrate presented two bands crossing E$_F$. However, this claim requires further investigation with films of different thicknesses and other substrates.

\subsection{Comparison to conflicting SARPES results\label{SOM:sarpes}}

Triggered by the SARPES results on STO crystals published in \cite{Santander:2014}, another group has attempted to reproduce these findings under different conditions, but found no clear spin polarization signal \cite{McKeown:2016}. This discrepancy deserves attention, although a full explanation will probably require both further studies and input from the authors of \cite{McKeown:2016}. First of all, one can consider the results presented here as a further verification of \cite{Santander:2014} for a slightly different system. In both cases the spin signal is clear in the raw data and requires no further analysis to determine the number of spin-polarized peaks. Also the obtained peak positions match well with those found for high resolution ARPES data, which further solidifies the conclusion. That the obtained spin polarization is not an artefact of the COPHEE end station follows from the large number of SARPES results from this machine which have been reproduced by other groups or by one-step photoemission theory. The combination of Rashba-like spin-orbit interaction and a Zeeman-like gap around the zone centre are the simplest explanation of the observed spin texture, but this does not exclude the possibility of a more complex explanation.

\begin{figure}[htb]

.	\includegraphics[width=0.8\columnwidth]{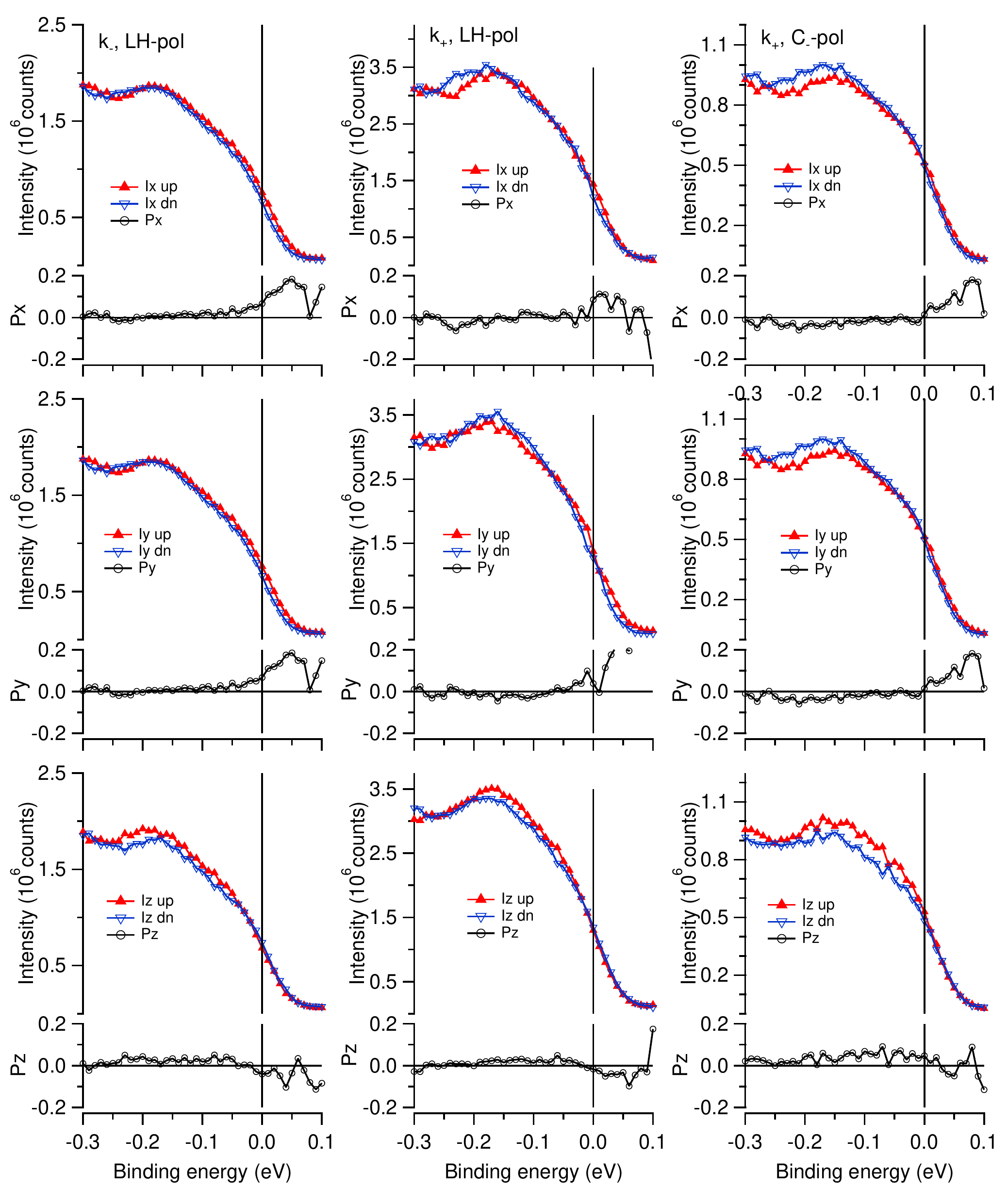}
	\caption{Spin-polarized EDCs of measured at COPHEE with 85 eV, using the same STO crystals as in \cite{Santander:2011}. k$_{+,-}$ denotes the Fermi momenta to the left and right of $\overline{\Gamma_{10}}$.}
	\label{fig.sarpes_comp}
\end{figure}

Now let us consider the differences in \cite{McKeown:2016} with respect to \cite{Santander:2014}. In \cite{McKeown:2016} the samples are La-doped and cleaved, but in our current understanding this should not make a large difference. The measurements in \cite{McKeown:2016} are performed on a different SARPES end station which uses a Mott detector that is less stable, as described in \cite{Petrov:2001}. However, also this should ideally have no large influence. The most importance difference, that can most likely explain the absence of a spin signal to a large extent, is the used photon energy. In \cite{McKeown:2016} a photon energy of 80~eV is used which coincides with the photon energy range where the $d_{yz,xz}$-derived heavy bands are strong and disperse to overlap with the $d_{xy}$-derived states for the crystal surface. Note that for the films these heavy bands are not occupied. Due to their strong 3D character these states show almost no spin polarization but they can mask the spin signal from the $d_{xy}$-derived states. This is also clear from the SARPES data shown in Fig.~\ref{fig.sarpes_comp} obtained at $h\nu=85$~eV at the COPHEE end station for the STO crystal surface. This data was obtained in 2012 on the exact same samples and during the same measurement run as in \cite{Santander:2014}. Clearly no significant spin polarization signal can be distinguished also in this case.

Lastly, there are some other points in \cite{McKeown:2016} that could explain the absence of a spin signal, but these would require more in-depth verification. Most importantly the data in Figure 2(b) of \cite{McKeown:2016}  appears to show a highly contaminated surface and is markedly different from the data in Figure 2(c) or what was observed in \cite{Santander:2014}. Whether under these conditions a spin signal can be expected requires further studies, but most likely the charge from adsorbates will significantly alter the properties of the 2DEG.

\subsection{Details of the XPS analysis\label{SOM:xps}}

\begin{figure}[htb!]
	\includegraphics[width=0.7\columnwidth]{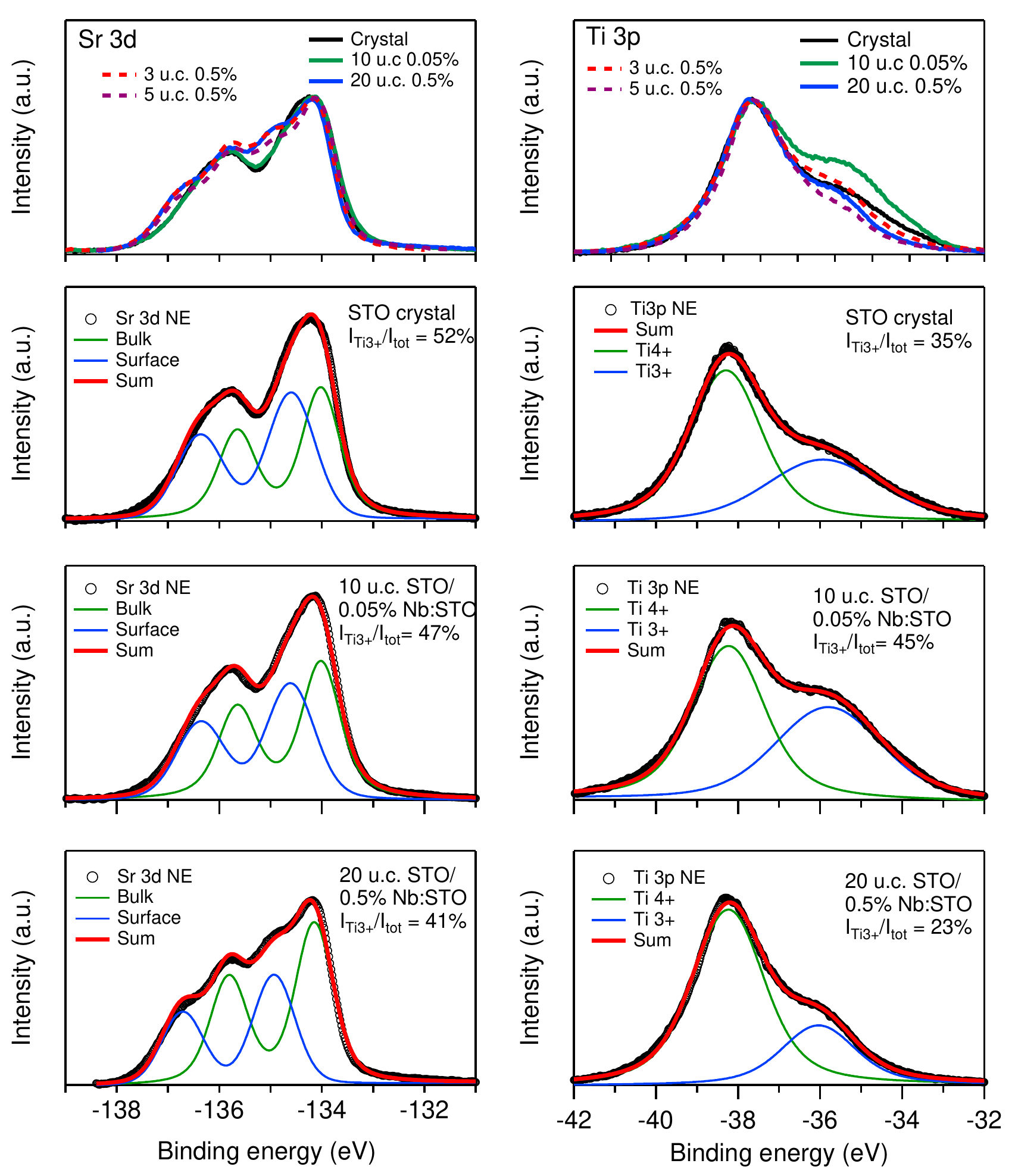}
	\caption{Details of XPS fits. (a) Sr 3d and (b) Ti~3p spectra of different samples (top panel) and their fits (lower panels). The spectra of the 3 and 5 u.c. films is shown for comparison in dashed lines.}
	\label{fig.xps_supp}
\end{figure}

The top panels of Fig.~\ref{fig.xps_supp} shows the Sr 3d and Ti 3p XPS spectra of the STO crystal, 20 u.c. films on 0.5\% Nb:STO and 10 u.c. films on 0.5\% Nb:STO, along with their respective fits in the lowe panes. For simplicity, we omitted the fits of spectra of the 3 and 5 u.c. films, which are only shown for comparison as dashed lines in the top panels. The XPS analysis done in the main text is based on the fits  shown in Fig.~\ref{fig.xps_supp}, where each of the peaks is a Lorentzian function convoluted with a Gaussian one. A Shirley-type background was subtracted from all spectra. 

For the Sr 3d spectra, we used two pairs of peaks (for bulk and suface components), each with a fixed ratio of 2/3 between the spin-orbit contributions. The average spin-orbit splitting of the bulk and surface Sr varied between 1.68~eV and 1.75~eV, with a variation smaller than 2\% among the spectra. The Sr/Ti ratio, calculated taking into account the Ti 3p and Sr 3d photoionization cross-sections \cite{Yeh:1985}, was found to be 1.4 in the studied crystal and varies between 1-1.20 in the studied films, as described in the main text. For the Ti 3p spectra, two independent peaks were used, representing the Ti$^{3+}$  and Ti$^{4+}$ contributions. The ratio of the surface and bulk contributions is shown along with the fits in the lower panels. For Sr 3d, a trend is observed: from STO crystal, going through the film on 0.05\% substrate and to the film on 0.5\% substrate, the amount of surface Sr systematically decreases. In turn, no clear trend is observed in the Ti 3p spectra with regard to the amount of Nb doping.

\bibliographystyle{apsrev4-1}
\bibliography{ReferencesFull}

\pagenumbering{gobble}

\end{document}